\documentclass[twocolumn,showpacs,prb,preprintnumbers,amsmath,amsfont,amssymb]{revtex4}
\usepackage{graphicx}
\usepackage{dcolumn}
\usepackage{bm}

\begin{document}

\title{Variation of the Density of States in Amorphous GdSi \\
at the Metal-Insulator Transition}
\author{L. Bokacheva$^a$} 
\author{W. Teizer$^b$} 
\author{F. Hellman$^a$} 
\author{R. C. Dynes$^a$}
\affiliation{$^a$Department of Physics, University of California, 
San Diego, La Jolla, CA 92093-0319 \\
$^b$Department of Physics, Texas A\&M University, 
College Station, TX 77843-4242}

\date{\today}

\begin{abstract}
We performed detailed conductivity and tunneling mesurements on the amorphous, 
magnetically doped material $\alpha$-Gd$_x$Si$_{1-x}$ (GdSi), which can be 
driven through the metal-insulator transition by the application 
of an external magnetic field. Conductivity increases linearly with field 
near the transition and slightly slower on the metallic side. 
The tunneling conductance, proportional to the density of states $N(E)$, 
undergoes a gradual change with increasing field, from insulating, 
showing a soft gap at low bias, with a slightly weaker than parabolic 
energy dependence, i.e. $N(E) \sim E^c$, $c \lesssim 2$, towards metallic 
behavior, with $E^d$, $0.5<d<1$ energy dependence. 
The density of states at the Fermi level appears to be zero at low fields, 
as in an insulator, while the sample shows already small, but metal-like 
conductivity. We suggest a possible explanation to the observed effect.

\end{abstract}
\maketitle

\section{\label{sec:Intro}Introduction}

The metal-insulator transition (MIT) in disordered systems 
\cite{MITReviews} 
is a complex problem, complete and accurate treatment of which requires 
that the effects of localization and the electronic correlations 
be taken into account simultaneously and on equal footing.
The main difficulty in constructing such a model lies in the fact 
that near the MIT strong Coulomb interactions cannot be treated 
perturbatively and thus render single-particle models unrealistic. 
Carriers that are localized by the disordered potential cannot screen 
the Coulomb interactions as well as the mobile ones, therefore in 
disordered systems the~correlations play an important role and must be 
included in the picture of the MIT.

Far from the transition, theoretical models have successfully described 
the effects due to the interactions in the conductivity 
$\sigma$ and density of states (DOS) $N(E)$.
The transport conductivity of a metallic material with strong Coulomb 
correlations as a function of temperature follows 
a power law: $\sigma(T)=\sigma_0+\sigma_1 T^y $, with $y=0.5$.
\cite{MegaAronMetal}
On the insulating side, the interactions modify the exponent of the temperature 
dependence of the variable range hopping (VRH) 
conductivity from $1/4$ in the noninteracting case \cite{Mott} to $1/2$:
$\sigma(T)\sim\exp(-(T_0/T)^{1/2})$. \cite{EfrosShklovskiiConduct} 
In the density of states, on either side of the MIT, interactions generally 
cause the depletion of states near the Fermi energy. 
In~the metallic regime the correlations are manifested as a~square-root dip 
in the DOS: $N(E)~=~N(0)(1+(E/\Delta)^{1/2})$, 
where $\Delta \sim \hbar D/l^2$ is the correlation energy, $D$ is 
the diffusion coefficient, $l$ is the mean free path, and $\hbar$ is Planck's 
constant. \cite{MegaAron,McMillan} 
In an insulator with thermally activated variable range hopping conductivity, 
the interactions lead to the~opening of a~soft gap with a quadratic 
(in 3D case) energy dependence, $N(E)=(3/\pi )(\kappa /e^2)^3E^2$, 
where $\kappa$ is the dielectric constant and $e$ is the electronic 
charge. 
The width of the Coulomb gap \cite{EfrosShklovskiiDOS} $\Delta_c$ is determined 
by $\kappa$ and the noninteracting density of states $N_0$: 
$\Delta_c=e^3(N_0/\kappa ^3)^{1/2}$.  

The presence of the features due to the Coulomb interactions in the 
conductivity and the DOS of disordered systems near the MIT has been well 
established experimentally.
The usual way to investigate the transition is to study a set of samples with 
different dopant concentrations near the critical one. \cite{Concentration} 
In these samples one can measure the transport conductivity and in some cases, 
the tunneling conductance, which is believed 
to be approximately proportional to the DOS. Features described above, 
such as VRH with $1/2$ exponent, the Coulomb gap in the DOS on the insulating 
side, the square root of T conductivity dependence and the square root cusp 
in the DOS have been observed in various experiments performed on
crystalline \cite{MarkLee} and amorphous samples. \cite{NbSi}

Amorphous systems have an advantage over the crystalline materials for the MIT 
studies, since they undergo the transition at doping concentrations that are 
orders of magnitude higher than the crystalline materials 
(roughly $x \sim 10^{-1}$ versus $x \sim 10^{-5}$).
As a consequence, the Fermi temperature $T_F$ of amorphous systems is much 
higher than in the crystalline ones, and $T/T_F$ is much lower. 
While it becomes inappropriate to think in terms of single particles, 
it is convenient to use the terminology to 
identify the conditions where these correlation effects dominate.
For amorphous systems the region where the Ioffe-Regel localization 
criterion \cite{IoffeRegel} $k_F l \sim 1$ is valid 
(here $k_F$ is the Fermi wave vector) is significantly expanded, and 
the maximum critical conductivity $\sigma =ne^2/(\hbar k_F^2$), where $n$ 
is the concentration, can reach 
$500$~($\Omega$\,cm)$^{-1}$ at $n\sim 10^{22}$ cm$^{-3}$, 
as opposed to $20$~($\Omega$\,cm)$^{-1}$ in the crystalline doped semiconductors. 
\cite{Concentration} 
Therefore, in amorphous materials one can probe the transition much deeper 
in the critical regime at accessible temperatures. 

In some materials it is possible to observe the MIT in a~single sample 
and tune it by varying an external parameter, such as magnetic field, 
\cite{MagField} stress, \cite{Stress} or illumination. \cite{Illumination} 
This approach is preferable to studying multiple samples, because 
it eliminates the undesirable scatter in the sample parameters, inevitable 
when using a discrete set of samples. 
For this reason, magnetically doped amorphous materials, such as 
$\alpha$-Gd$_x$Si$_{1-x}$ (GdSi) and Tb$_x$Si$_{1-x}$, \cite{PRL1,Tb} are
particularly suited for the studies of the MIT. 
In these materials besides the structural disorder, there is 
an additional degree of disorder associated with random orientation of the 
ionic magnetic moments, which can be controlled by the magnetic field. 
When the magnetic moments of the impurity ions are aligned by the external 
field, the disorder in the system
is reduced, and the mobility edge $E_c$ is lowered. \cite{PRL} 
This promotes delocalization of carriers and increases the conductivity 
of the material. 
Therefore a single sample with the concentration slightly below critical 
and thus insulating can be continuously 
driven through the MIT into the metallic state by an external 
magnetic field. It is also possible to probe the density of states (DOS) 
simultaneously with the transport properties of these materials by studying 
the tunneling of quasiparticles through a thin oxide barrier 
between the material in question and a metallic electrode. 
These experiments enable one to track the changes that occur in the DOS 
at the transition and to separate them from 
the mobility effects, which both contribute to the conductivity.

Insulating and metallic behavior can be defined as follows. 
The sample is an insulator, if its conductivity 
$\sigma(T)$ extrapolated to zero temperature vanishes: $\sigma(T~=~0)~=~0$. 
If $\sigma(T=0)$ is finite, the sample is on the metallic side 
of the transition. 
In GdSi the MIT occurs at $x \approx 0.14$ of Gd in zero magnetic field. 
Near this critical doping value, $\sigma(T=0)$ is very sensitive 
to the slight variations of $x$. 
A sample doped slightly below the critical level at low temperature 
shows a large positive 
magnetoconductivity and can be continuously driven through the transition 
by the magnetic field. \cite{SolStateCommun} 
Near the MIT the amorphous GdSi has been extensively studied: 
its structure, \cite{XAFS} transport conductivity, 
\cite{PRBRapid,PRL1,SolStateCommun} 
tunneling conductance, \cite{PRL,Proceedings} 
specific heat, \cite{HeatCapacity} magnetization, \cite{Magnetization} 
infrared absorption, \cite{Basov}
spin polarization, \cite{SpinPolarization} and 
the Hall effect, \cite{Hall} have been investigated in detail. 
It has been shown that the magnetic field plays a role similar to 
increasing the Gd concentration $x$ and increases the conductivity while 
raising both the free carrier concentration and the density of states, 
which appear to grow towards the metallic side 
in a coordinated way. \cite{PRL} In further work the density of free 
carriers has been found to vary linearly with magnetic field. \cite{Hall}

The results of tunneling measurements \cite{PRL} indicate that 
deep in the metallic regime, yet at $\sigma_0$ still below 
the Ioffe-Regel condition, the DOS of GdSi has $E^{1/2}$ 
dependence with a non-vanishing $N(0)$. 
Tunneling conductance curves $dI/dV(V)$ at different constant magnetic 
fields are approximately parallel to each other 
with only $N(0)$ increasing with field. 
On the insulating side these experiments are limited by the increasing 
resistance of the GdSi film. When the
resistance of the insulating GdSi film over the area of the tunnel 
junction becomes comparable with the resistance 
of the junction itself, the voltage drop occurs both over the junction 
and the film, so that the bias voltage cannot 
be considered a true energy coordinate. 
In samples which become conducting in magnetic field, the DOS shows a soft 
gap near zero bias with somewhat 
weaker than $E^2$ dependence and $N(0)$ vanishing at low fields. 
The tunneling data also suggest that in the transition region the DOS 
evolves continuously from insulating, showing a soft gap at zero bias, 
to metallic, with a square root cusp, upon increasing magnetic field. 
Near the MIT the DOS at the Fermi level $N(0)$ has been shown to depend 
on magnetic field as $(H-H_c)^2$. \cite{PRL} 
The models for the DOS described above that are valid far from 
the transition, cannot be applied in the critical region 
and cannot be properly merged at the critical concentration. 
It is not known how the DOS evolves from the insulating form to 
the metallic one. 
It~is therefore highly desirable to be able to tune a single sample 
through the MIT and measure its transport and 
tunneling properties as close to the transition as possible.
 
\begin{figure}
\includegraphics[trim=90 20 100 20]{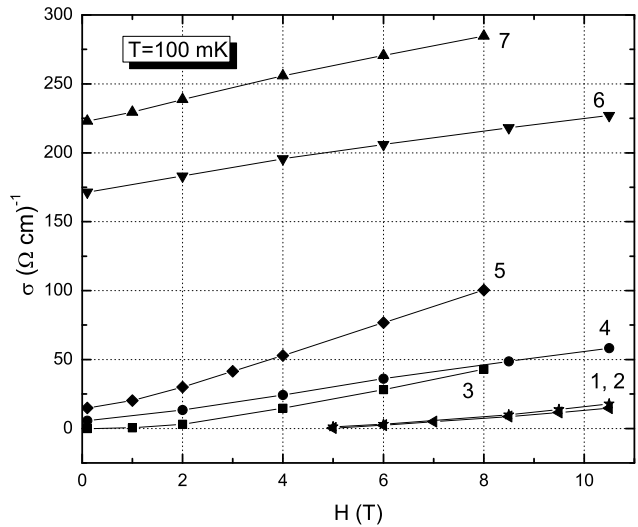}
\caption{\label{fig:CondvsH} Low temperature conductivity 
$\sigma (H)$ of seven different samples versus the applied magnetic field. 
At $H=0.1$~T samples \#1 and \#2 are insulating, sample \#3 is at 
the transition, samples \#4 and \#5 are slightly metallic, 
samples \#6 and \#7 are in the metallic regime.}
\end{figure}

\section{\label{sec:Experiment} Experimental Results}

In this paper we focus on the region in the immediate vicinity of 
the MIT and investigate the simultaneous 
variation of the transport conductivity and the DOS in a~detailed 
and systematic way. 
The results presented here were obtained using GdSi samples prepared 
as described previously. \cite{PRL} 
Amorphous films of GdSi, $100$~nm thick, have been grown by e-beam 
co-evaporation on Si/SiN substrates. On top of the GdSi film
tunnel junctions were formed with a thermally grown oxide barrier 
and Pb counter electrodes. 
The tunnel junctions were also used as voltage terminals in the four 
point transport measurements. 
We selected only those samples, for which the junction resistance 
was at least an order of magnitude higher than the film resistance 
over the area of the junction.

\begin{figure}[t]
\centering
\includegraphics[trim=90 20 100 20]{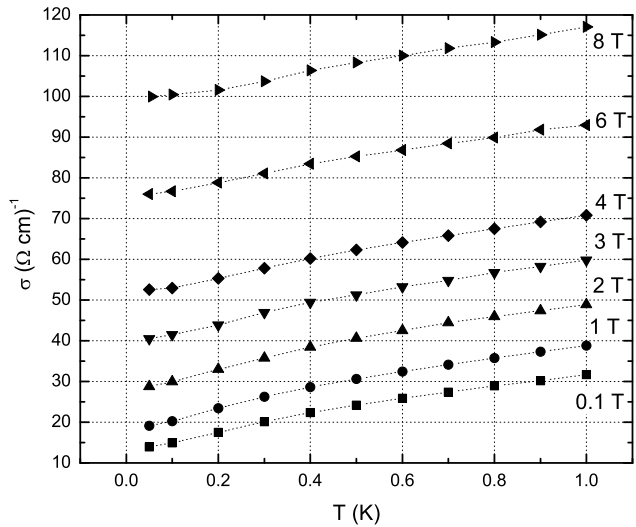}
\caption{\label{fig:CondvsT} Conductivity of a slightly metallic GdSi 
sample (sample \#5) versus temperature.}
\end{figure}

\subsection{\label{sec:Transport} Transport conductivity measurements}

At low temperatures (below approximately $50$~K) GdSi shows a large 
positive magnetoconductance. \cite{PRBRapid} 
In order to determine the sample's proximity 
to the MIT and to avoid the ambiguity associated with the extrapolation 
of $\sigma(T)$ to $T=0$, we use the actual conductivity 
values at the lowest 
temperatures available in our experiments ($\leq 100$~mK). 
The relative increase of the conductivity with field is the largest 
in the insulating samples, 
where the conductivity grows as a faster than linear function of 
the applied field. 
In samples which are right at the transition or barely on the metallic 
side, the conductivity at a constant low temperature 
increases approximately linearly with magnetic field. 
Far on the metallic side this dependence is slightly weaker than linear. 
Fig.~\ref{fig:CondvsH} shows the low temperature ($T=100$~mK) conductivity 
versus the applied magnetic field for 
several different samples, with the conductivity 
spanning a broad range from zero to over $200$~($\Omega$\,cm)$^{-1}$, all below 
the Ioffe-Regel limit. 

The conductivity of the most insulating samples \#1 and \#2 
becomes measurable above $4-5$ T and grows rapidly with field from 
nearly zero ($0.4$~($\Omega$\,cm)$^{-1}$) to 
$14-17$~($\Omega$\,cm)$^{-1}$ at $H=10.5$~T, i.e., between the lowest and 
the highest field the conductivity in these samples 
increases by $40$ or $50$ times. 
Sample \#3 is very close to the transition, and samples \#4 and \#5 are 
barely on the metallic side of the MIT. 
The conductivity of samples \#3, \#4 and \#5 is approximately linear 
with field. In sample \#3 the conductivity grows 
between $0.1$~T and $8$~T by a~factor of 40. The increase in the 
conductivity becomes much smaller 
for slightly metallic samples \#4 and\#5, in which it increases 
by a factor of 10 and 7 respectively.
Conductivity of the most metallic samples \#6 and \#7 is a slightly 
sublinear function of the field ($\sim H^{0.85}$) and in the shown range 
of field increases only by about 25\% of its value 
at $H=0.1$~T.

\begin{figure}[t]
\centering
\includegraphics[trim=70 20 80 20]{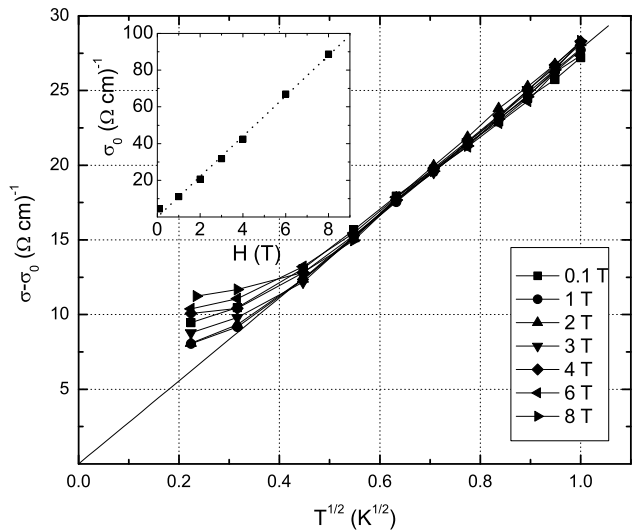} 
\caption{\label{fig:SigmavsSqrtT} Conductivity $\sigma-\sigma_0$ from 
the data on Fig.~\ref{fig:CondvsT} versus $T^{1/2}$ with $\sigma _0$ 
determined from linear fits to $\sigma ((T^{1/2}))$ subtracted. 
Solid line is a~linear fit to the data. 
Inset: $\sigma_0$ vs magnetic field $H$. Dotted line is a~linear fit.}
\end{figure}
  
Fig.~\ref{fig:CondvsT} shows the conductivity versus temperature 
curves for the barely metallic sample \#5 
($\sigma(50$~mK$)~=~14$~($\Omega$\,cm)$^{-1}$)  
obtained from the dc $IV$ transport measurements at constant fields 
from $0.1$~T to $8$~T. 
As shown before, \cite{PRL} the~conductivity curves can be fit with 
the power law $\sigma(T)= \sigma_0+\sigma_1T^y$ with the exponent 
$y \approx 0.6-0.7$, 
slightly higher than the predicted $y=0.5$ for a metal with correlations. 
The curves at different fields are approximately parallel to each other, 
only shifted towards higher values, 
which implies that $\sigma _0$ grows roughly linearly with field, 
as illustrated on Fig.~\ref{fig:CondvsH}, and $\sigma _1$ 
is weakly dependent on $H$. 
This can be easily seen in Fig.~\ref{fig:SigmavsSqrtT}, which shows 
the same data as in Fig.~\ref{fig:CondvsT}, plotted as a~function of $T^{1/2}$. 
The~lowest temperature offset $\sigma _0$, in this case determined 
from the linear fits to $\sigma(T^{1/2})$, has been subtracted from 
each curve. The linear variation of $\sigma_0$ with field is shown 
in the inset of Fig.~\ref{fig:SigmavsSqrtT}.
The $\sigma(T^{1/2})$ data at different fields overlap well and except 
at the lowest temperatures, fall onto a straight line. 
This dependence has been observed in non-magnetic materials, which 
supports the idea that the main effect of the applied
magnetic field is to decrease the disorder and lower the mobility 
edge, thus increasing the conductivity and driving the sample further 
into the metallic regime.

\subsection{\label{sec:Tunneling} Tunneling conductance measurements}

\begin{figure}[t]
\includegraphics[trim=75 20 80 20]{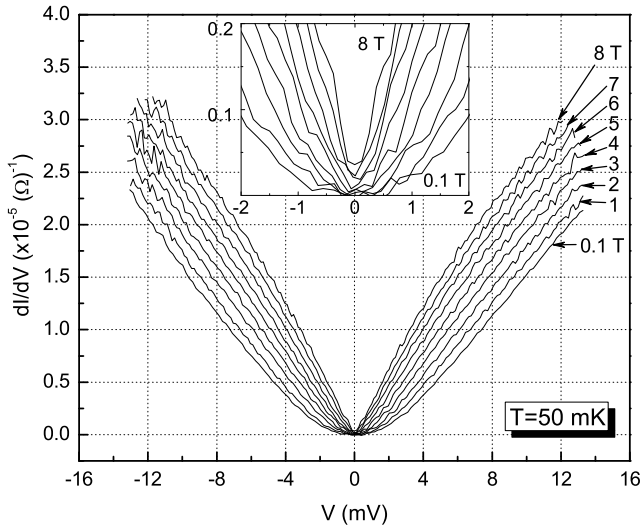}
\caption{\label{fig:dIdV} Tunneling conductance $dI/dV(V)$ curves at 
$T=~50$~mK and different magnetic fields. 
Inset: Same data in the low bias region. The axes are in the same units 
as the main figure.}
\end{figure}

To elucidate the crossover in the DOS from the insulating to the metallic 
behavior we performed detailed 
tunneling studies. The data presented below are the results of the dc 
$IV$ measurements. 
The dc technique is more appropriate in this case because at low 
temperatures near the MIT the resistance and the dielectric constant 
of GdSi films are strongly frequency dependent. 
The $IV$ curves measured at constant magnetic fields and a constant 
low temperature were numerically differentiated in order to obtain 
the differential tunneling conductance $dI/dV(V)$. 
Fig.~\ref{fig:dIdV} shows a set of such curves for sample \#5 at $50$~mK 
and magnetic fields from $0.1$~T to $8$~T. 

At the lowest field $H=0.1$~T (required to quench the superconductivity 
of the Pb counter electrodes) the tunneling conductance curve 
is superlinear in the whole range of the bias voltage 
between $\pm 12$~mV, but cannot be fit with a simple function. 
Near zero bias at $|V|<2.5$~mV the curve can be fit with a power law 
$a+b|V|^c$, with $c\sim1.72$. 
Outside this region the exponent $c$ is only slightly greater than unity: 
$c\sim1.1$. 
At $V=0$ the curve touches the horizontal axis, i.e. zero bias 
conductance is unmeasurably small: $dI/dV(V=0)=0$. 
Upon increasing the field, the curves undergo a gradual and continuous 
change. 
The~gap region with $c\agt 1.5$ contracts and by $8$~T it exists only 
in a small interval at $|V|<0.6$~mV. 
The~part of the~curve at higher bias expands and at the same time 
the~exponent $c$ in this region decreases 
from nearly unity at $0.1$~T to $0.74$ at $8$~T, apparently approaching 
the $V^{1/2}$ behavior seen
well into the metallic regime. 
Starting at about $H=2$ T, the $dI/dV$ curves have an inflection point, 
separating the two parts of the curve with $c>1$ 
(positive curvature), implying a~``soft gap'' and $c<1$ 
(negative curvature). 
At the highest field $H=8$~T, the $dI/dV$ curve is sublinear almost 
in the entire range of the bias voltage except 
a small region of $|V|<0.6$~mV near zero, and is characteristic 
of the metallic regime. 

\begin{figure}[t]
\includegraphics[width=1.95in,height=2.75in,trim=70 20 65 20]{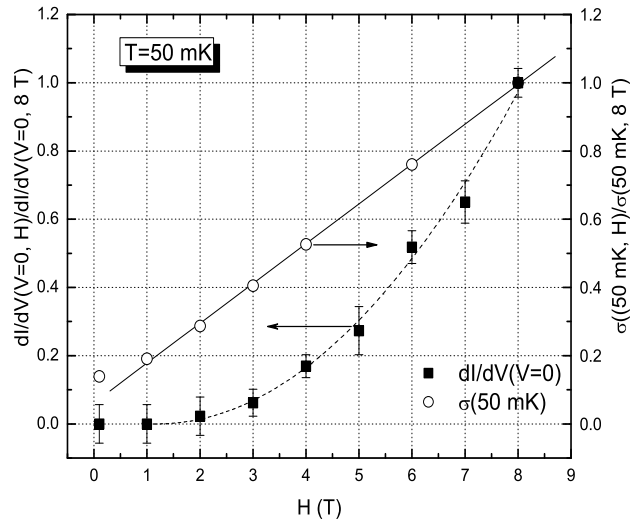}
\caption{\label{fig:ZBC} Zero bias conductance $dI/dV(V=0)$ 
(full squares) from the data on Fig.~\ref{fig:dIdV} and 
conductivity at $T=50$~mK (empty circles), both normalized to 
their respective values at $H=8$~T, versus magnetic field. 
Dashed line is the fit of $dI/dV(V=0)$ with  
$0.02 \times (H-1.2)^2$. Solid line is a linear fit to $\sigma_0(H)$.}
\end{figure}

This continuous crossover can be seen in more detail in the inset 
of Fig.~\ref{fig:dIdV}, which shows a zoom-in of 
the data in the low bias region below $2$ mV. 
Zero bias conductance appears to be zero at $0.1$~T and $H=1$~T, 
but increases at higher fields. 
In Fig.~\ref{fig:ZBC} zero bias conductance, normalized to its value at $8$~T, 
is plotted versus $H$ and can be well fit 
with $(H-H_c)^2$ with $H_c\approx1.2$~T. 
Note that while $dI/dV(V=0)\sim N(0)\approx 0$ at the two lowest 
fields, $0.1$~T and $1$~T, the sample has already an appreciable 
transport conductivity, 
$14$ and $19$ ($\Omega$\,cm)$^{-1}$ respectively (see also Fig.~\ref{fig:CondvsT}). 
At these fields the sample exhibits metallic transport properties, 
but appears to have an unmeasurably small number of  
extended states at the Fermi level from the tunneling experiments. 

\begin{figure}[t]
\includegraphics[width=3in,trim=30 20 40 20]{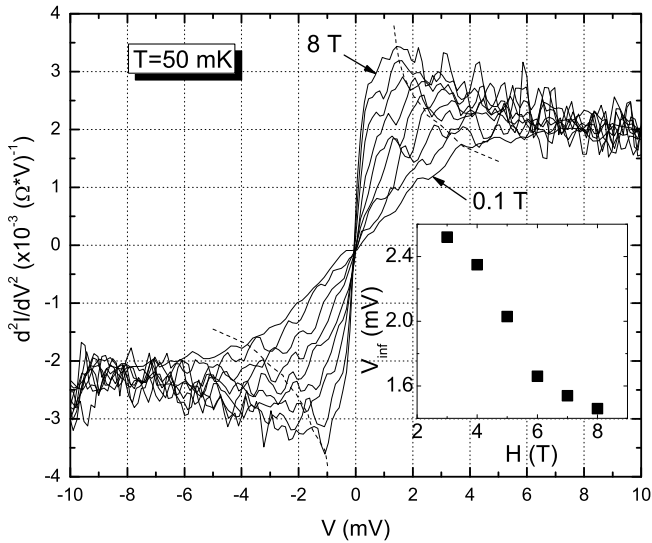}
\caption{\label{fig:SecDeriv} Second derivative of the $IV$, 
$d^2I/dV^2(V)$ for the data on Fig.~\ref{fig:dIdV}. 
Dashed line is a guide for the eye indicating the position 
of a maximum shifting to lower bias with increasing field. 
Inset: The bias voltage $V_{\rm inf}$ at which the inflection 
point in the $dI/dV$ curves is located, plotted versus magnetic field.}
\end{figure}

The gradual change in the shape of the $dI/dV(V)$ curves
from an insulating $V^2$ to a metallic $V^{1/2}$ dependence 
and the position of the inflection point can be easily seen 
on a plot of the second derivative of the $IV$ curves, 
$d^2I/dV^2(V)$ shown on Fig.~\ref{fig:SecDeriv}. 
The $d^2I/dV^2$ curves reflect the curvature of the $IV$ 
and the slope of the $dI/dV$ curves. 
Clearly, the curvature is the largest and positive in the region 
near zero bias.
With increasing field this region shrinks, and the point of inflection 
moves to lower voltage values. 
The maximum in $d^2I/dV^2$ corresponds to the change of curvature 
in the $dI/dV$ from positive to negative. 
In order to determine the position of the inflection point in $dI/dV$, 
we take a~maximum of the $d^2I/dV^2$ 
on Fig.~\ref{fig:SecDeriv}. 
The~curves at the lowest fields, $0.1$ T and $1$~T, do not have a~maximum, 
although their slope significantly decreases 
at $3.75$~mV and $3.6$~mV respectively and falls almost to zero 
above $6$~mV. 
Upon increasing the field, a maximum appears in the $d^2I/dV^2$ curve, 
which corresponds to an inflection point in $dI/dV$, and moves towards 
$V=0$. The bias voltage, at which the inflection occurs, decreases with 
increasing field, as shown in the inset of Fig.~\ref{fig:SecDeriv}.


The significance of these data, we believe, is the indication that 
the crossover from metallic $V^{1/2}$ behavior 
to insulating $V^2$ behavior is continuous and we cannot identify from 
the tunneling data a ``critical point'' that marks 
the transition. 

We emphasize that this study probes the sample very 
close to the MIT with high resolution. 
This is achieved for two reasons. Firstly, the transition is continuously 
tuned with applied field, and secondly, the region
of study is well below the Ioffe-Regel condition. 
We have observed similar behavior in the tunneling characteristics of 
all samples studied to 
date in the vicinity of the MIT. In fact, the $dI/dV(V)$ data at 
different fields for different samples can be scaled to form 
one ``master set'' of curves, 
progressively changing their shape from insulating to metallic, 
correlated with the sample conductivity.
 
The results of the tunneling experiments can be summarized as follows. 
The DOS evolves gradually and continuously with increasing magnetic 
field from an insulating to a metallic shape. 
The Coulomb correlation gap, associated with the insulating behavior, 
shrinks as the field increases. 
At these fields, transport experiments on the sample already show 
metallic transport conductivity. 
Zero bias tunneling conductance grows as a strong (quadratic) function 
of the applied field. 
For samples very close to the transition, one can observe a situation, 
where the tunneling conductance is vanishingly small, 
and the transport conductivity is already appreciable. This is a result 
we did not expect, as we anticipated
that $N(E)$ and $\sigma_0$ would vanish together at the transition.
We conclude that the effects in the density of states $N(E)$ that are 
associated with 
the insulating regime persist well into the metallic regime.

\section{\label{sec:Picture} Discussion}

Observation of a measurable metallic conductivity simultaneously with 
an apparent absence of the extended states 
at the Fermi level appears paradoxical at first sight. 
Nevertheless, a situation where this is possible can be easily pictured. 
Consider a strongly disordered system approaching the MIT from 
an insulating side. 
When at the transition a small fraction of all carriers becomes 
delocalized, the current may flow via a very small number of conduction 
paths among the disordered sites. 
Conductivity due to these paths is much higher than that due to the rest 
of the carriers, which are still localized 
and participate in transport by the thermally activated variable-range hopping. 
At low temperature the contribution of the localized carriers to
the conductivity is exponentially small compared to the contribution 
of the extended states, despite the fact that the former
constitute an overwhelming majority among all carriers. 
The overall conductivity therefore will be determined by the few, 
high mobility extended carriers. 
On the other hand, the DOS that is observed in an experiment, 
such as ours, reflects an average number among 
all microscopic states in the sample at a given energy, thus its 
magnitude will be determined by the dominating species. 
Near the transition a small number of carriers possess a metallic DOS 
with $N(E)\sim E^{1/2}$ and $N(0)\neq 0$, however
a vast majority of states still show an insulating type of DOS with 
a parabolic energy dependence and a vanishing $N(0)$. 
When summed over all microscopic states with appropriate weights 
proportional to the numbers of each species, 
the contribution to the DOS of the localized states overwhelmingly 
prevails over that of the metallic ones, and as a result, 
at the transition the total DOS appears to be insulating. 
In this case the DOS at $E=0$ at the MIT is likely to vanish.
As the sample is driven towards the metallic regime and the number 
of the extended carriers grows, the DOS acquires 
an intermediate shape, i.e., it shows features due to both insulating 
and metallic species simultaneously. 
Gradually the signatures of the metallic behavior begin to take over 
larger and larger range of energy, until eventually 
they dominate the whole DOS. At the same time, while the sample is 
tuned into the metallic regime, its conductivity rapidly increases 
and exhibits metallic behavior.   
Therefore, it is plausible that in the same conditions, the transport 
experiments will place this sample 
on the metallic side of the transition, while in the tunneling 
measurements it will appear as an insulator.

It could be argued that this description is a percolation transition 
on an atomic scale and that a percolation threshold 
in a two component system in 3 dimensions should occur at a fractional 
concentration of $x_c\sim0.43$ (for site percolation 
in a diamond structure). \cite{Percolation} 
However
this transition must take into account in addition to the natural 
inhomogeneity at the atomic scale, the strong correlations
that are ignored in a classical model. Indeed, while we resort to 
a single particle picture of mobility and density
of states for descriptive reasons, we must remind ourselves that 
this transition is believed to be driven by the correlations
which are not single particle effects. We believe that these current 
experiments show that thinking in terms of independent 
particles in the vicinity of the metal-insulator transition, 
despite the convenience of using such pictures, is an oversimplification. 
Further theoretical work may be able to put these experimental results 
on a solid footing within the correlation-based models.

\begin{acknowledgements}
We thank S. Kivelson, O. Naaman, and A. Frydman for valuable 
discussions. This work was supported by the NSF 
Grant No. DMR-0097242.
\end{acknowledgements}

\end{document}